\documentclass[preprint,prd,amssymb,amsmath,nobibnotes,nofootinbib]{revtex4} 
\usepackage{amsfonts} 
\usepackage{revsymb} 
\usepackage{graphicx,epsfig} 
\usepackage{placeins} 
\begin{document} 
\title {Regularization of second-order scalar perturbation 
 produced by a point-particle with a nonlinear coupling} 
\author{Eran Rosenthal}
\address{Department of Physics, University of Guelph, Guelph, Ontario  N1G 2W1, 
Canada}

\date{\today} 
 
\begin{abstract} 
Accurate calculation of the motion of a compact object 
in a background spacetime induced by a supermassive black hole
is required for the future detection of such binary systems 
 by the gravitational-wave detector LISA. 
Reaching the desired accuracy requires calculation of the 
second-order gravitational perturbations produced by the compact object. 
At the point particle limit the second-order gravitational perturbation 
equations 
turn out to have highly singular source terms, for which the standard
retarded solutions diverge. Here we study a simplified scalar toy-model 
 in which a point particle induces a nonlinear scalar field in a given curved spacetime.
 The corresponding second-order scalar perturbation
 equation in this model is found to have a similar singular source term, and therefore its
  standard retarded solutions diverge.
We develop a regularization method for constructing well-defined causal
 solutions for this equation. Notably these solutions  
differ from the standard retarded solutions, which are ill-defined in this case. 
\end{abstract} 
 
\date{\today } 
 
\maketitle 
 
\section{Introduction} 
 
One of the interesting sources of gravitational waves which could 
be detected by the planned laser interferometer space antenna (LISA) \cite{LISA} 
is composed of a compact stellar object with a mass $\mu$ 
 (e.g., a neutron star or a small black hole) which inspirals towards 
 a supermassive black hole with mass $M$, such that $M\gg\mu$. 
Successful detection of these sources using matched-filtering data-analysis  
techniques requires preparation of gravitational  
waveform templates for the expected gravitational waves. 
The calculation of a complete waveform template corresponding
 to an inspiral orbit which lasts for one year, requires 
calculation of about $10^5$ wave cycles \cite{TC}. 
One of the main difficulties in the construction of templates for  
such long wave-trains is to keep track of the gravitational-wave accumulating 
phase. Accurate calculation of this phase requires 
accurate calculation of the compact object inspiral trajectory, which follows 
from the solution to Einstein field's equations for this problem. 
 
In practice one can take advantage of the smallness of the ratio $\mu/M$ and 
use perturbation analysis to simplify the calculation of the inspiral trajectory. 
For concreteness suppose that the small compact object is 
 a small Schwarzschild black hole, 
 which moves in a vacuum background spacetime induced 
by a supermassive black hole. 
Here, far from the small black hole (i.e., at distances much larger then $\mu$)  
the geometry is dominated by the supermassive black hole. In this region it
 is useful to express the full spacetime metric ${\sf g}_{\mu\nu}$ as a sum of 
a background metric  $g_{\mu\nu}$, induced 
by the supermassive black hole, and a sequence of gravitational 
perturbations that are induced by the small black hole, reading 
\begin{equation}\label{metricexpn} 
{\sf g}_{\mu\nu}(x)=g_{\mu\nu}(x)+\mu h^{(1)}_{\mu\nu}(x)+ \mu^2 h^{(2)}_{\mu\nu}(x)+O(\mu^3)\,. 
\end{equation} 
Here the perturbations $\{h^{(i)}_{\mu\nu}\}$ are defined on the background geometry,  
and are independent of $\mu$. 
By substituting this asymptotic expansion into Einstein's field equations one obtains linear  
field equations for the first-order and second-order gravitational perturbations 
 $h^{(1)}_{\mu\nu}$  and $h^{(2)}_{\mu\nu}$, respectively.  

To further simplify the calculation  
it is useful to consider the limit $\mu\rightarrow 0$ of the series (\ref{metricexpn}). 
Notice that by definition  $h^{(1)}_{\mu\nu}$ and $h^{(2)}_{\mu\nu}$ do not  
depend on $\mu$ and therefore the form of the perturbations equations which they satisfy is not  
affected by this limit. However, the domain of validity of these equation 
is in fact expanded as $\mu \rightarrow 0$.  
For a finite value of $\mu$ the domain of validity of  
expansion (\ref{metricexpn}) lies outside a worldtube  
of radius $r_E(\mu)$ such that $r_E\gg\mu$.
 At the limit $\mu\rightarrow 0$ the radius $r_E$ approaches zero and the series  
(\ref{metricexpn}) is valid throughout the entire background spacetime 
 excluding a worldline $\gamma$. 
Here D'Eath has shown \cite{DEATH} that in this limit the  
first-order metric perturbations $h^{(1)}_{\mu\nu}$ (in the Lorenz gauge) are identical to  
the corresponding perturbations that are produced by a unit-mass point particle tracing  
the same worldline $\gamma$. Therefore, for practical purposes
 (e.g., to calculate the emitted
gravitational waves) we can identify $\gamma$ as the worldline of  
the small black hole in the limit $\mu \rightarrow 0$.
 Note that for a finite value of $\mu$ this identification is 
valid only at distances from the small black hole which are much larger then $\mu$. 
 
Complementing the metric perturbations series, we also treat the worldline $\gamma$ 
perturbatively. 
Here at the leading order of approximation the worldline $\gamma$ is dominated  
 by the background geometry. In fact it was found to be a geodesic  
in the background spacetime (see e.g., \cite{DEATH}). 

At the next order, the interaction between the small black hole's mass $\mu$ 
and the gravitational field induced by the small black hole itself, 
gives rise to a gravitational self-force acting on the small black hole. 
The leading order effect of the self-force originates  
from interaction with the first-order gravitational perturbations $\mu h^{(1)}_{\mu\nu}$. 
This leading self-force (of order $\mu^2$) produces an acceleration of order $\mu$ 
for the black hole's worldline $\gamma$. For a vacuum background geometry, 
 formal and general expression for the leading order self-force on a small compact  
object was derived by Mino, Sasaki, and Tanaka \cite{MST},  
 and independently by Quinn and Wald \cite{QW} using a different method. 
Later, practical  
methods to calculate this self-force were developed by several authors 
 \cite{PP,BO1,BMNOS,BO2,BO3,LOUSTO,BL}; 
 see also \cite{MINO} for a different approach to this problem. 
 
The next order correction to $\gamma$ in this approximation scheme
originates from the interaction  
of the black hole's mass with the second-order gravitational perturbations 
 $\mu^2 h^{(2)}_{\mu\nu}$. 
Here it was shown by Poisson \cite{POISSON} by simplifying an argument by Burko   
\cite{BURKO} that the shift in the inspiral orbit   
due to this interaction may give rise to a shift in the gravitational-wave phase 
 of order $\mu^0$. This $O(1)$ shift in the phase is important for the  
construction of long and accurate waveform templates for LISA, and provides  
a practical motivation for the study of the second-order gravitational  
perturbations. 
 
Schematically (and without indices) the second-order perturbation equations,  
which are obtained from Einstein's field equations in the limit $\mu\rightarrow 0$, read
\begin{equation}\label{2ndgrav} 
D[h^{(2)}]=S \ \ \ ,x\not\in\gamma\,. 
\end{equation} 
Here $D$ denotes a linear differential operator, 
 and $S$ denotes the corresponding source term. 
Recall that in the limit of interest  
we may consider $h^{(1)}$ as 
being induced by a unit mass point-particle. Therefore, in the Lorenz gauge this field  
satisfies a linear wave equation with a point-particle source term.  
The solution of this wave equation  
has a well known $\epsilon^{-1}$ divergent behavior  
in the vicinity of the worldline $\gamma$, 
where $\epsilon$ denotes the spatial distance from $\gamma$. (Throughout this paper 
we shall define the spatial distance $\epsilon$ of a point $x$ from a worldline 
to be the length of a spacelike geodesic which connects the worldline with $x$.
 This geodesic is defined such that its tangent vector is normal to the worldline,
and it lies within a normal neighborhood of this worldline). 
Examining the form of the source term $S$ 
reveals that it contains two types of terms that can be schematically expressed as   
 $\nabla h^{(1)}\nabla h^{(1)}$ 
 and $h^{(1)}\nabla\nabla h^{(1)}$, 
where $\nabla$ schematically denotes the covariant derivative with respect 
 to the background metric. 
Here there is a problem, since the $O(\epsilon^{-1})$ divergent 
 behavior of $h^{(1)}$  
implies that source term $S$ diverges like $\epsilon^{-4}$. 
 A naive attempt to construct a standard retarded solution to Eq. (\ref{2ndgrav}), say by  
imposing Lorenz gauge conditions on $h^{(2)}$ and then formally integrating  
$S$ with the corresponding retarded Green's function,  
ends up with an ill-defined integral for the retarded solution which diverges at 
 every point in spacetime.

In this article we lay the foundations for a regularization 
method which can be generalized to enable one to construct a well-defined solution to
 Eq. (\ref{2ndgrav}). 
The construction of a well-defined solution is a first step towards 
 the calculation of the previously mentioned $O(1)$
 corrections to the gravitational-wave phase-shift, 
 which are produced by the interaction between the second-order gravitational perturbations 
and the small black hole. 
 
In this paper we study a simpler nonlinear scalar toy-model, which  
exhibit similar difficulties.
We consider a point-particle with a scalar charge $q$ which
induces a scalar field $\phi$ that 
satisfies a certain nonlinear field equation in a given curved background spacetime. 
Recall that in the corresponding gravitational problem we found it 
useful to consider 
the limit $\mu\rightarrow 0$. Similarly in our scalar model 
we shall seek a solution to the nonlinear scalar field
equation in the limit $q \rightarrow 0$. For small values of $q$ 
 we shall derive linear partial-differential-equations for 
 the first-order and second-order scalar perturbations. 
In the limit the source term in the second-order scalar perturbation equation 
turns out to be highly singular, and diverges like 
 $\epsilon^{-4}$.
 This singularity is similar to the   
 $\epsilon^{-4}$ singularity encountered in the source   
term of the corresponding second-order gravitational perturbations equation (\ref{2ndgrav}). 
Similar to the gravitational case,
 the standard retarded solution of the 
second-order scalar perturbation equation diverges at every point in spacetime. 
To resolve this difficulty,  
we develop a new regularization method. Essentially this method is based on 
constructing a new well-defined causal solution to  
the problematic second-order scalar equation. Notably this solution is  
different from the standard retarded solution which is ill-defined for this problem. 
The generalization of the regularization method presented here  
to the more physical, and more difficult second-order gravitational 
 perturbations will be published elsewhere. 
 
This article is organized as follows: First, in Sec. \ref{nonlinearfield} we
present our nonlinear scalar toy-model, and construct the relevant 
perturbations equations; then in Sec. \ref{firstorder} we 
construct the first-order scalar perturbation in the limit $q\rightarrow 0$;
and finally in Sec. \ref{secondorder} we consider 
the second-order scalar perturbation equation, for which we 
construct well-defined causal solution in the limit $q\rightarrow 0$.

\section {Nonlinear scalar toy-model} \label{nonlinearfield}
 
Consider a particle with a scalar charge $q$ which traces a geodesic 
 \footnote{Note that the problematic $O(\epsilon^{-4})$ source term in the corresponding 
gravitational equation (\ref{2ndgrav}) 
originates from terms involving $h^{(1)}$, where 
$h^{(1)}$ is induced by a unit mass which traces a geodesic worldline.
This motivates choosing a geodesic worldline for the scalar model as well. 
To keep our scalar particle on a geodesic trajectory we assume that external forces are 
acting on the particle such that the sum of all
 the forces which act on the particle vanishes --  
this sum of forces includes the scalar self-force and all the external forces.} 
 $z(\tau)$ in  
a given background spacetime. Here $\tau$ denotes the particle's proper time, 
with respect to the background geometry, the background metric is denoted with $g_{\mu\nu}$, 
and we adopt the signature $(-,+,+,+)$ throughout. 
In our model, the scalar field $\phi(x)$ satisfies 
 the following nonlinear field equation 
\begin{equation}\label{scalarfeq} 
\Box\phi = -4\pi q n+ \nabla_{\mu}\phi\nabla^{\mu}\phi\,. 
\end{equation} 
Here $\Box$ denotes the linear wave operator 
 given by $\Box\phi\equiv g^{\mu\nu}\nabla_{\mu}\nabla_{\nu}\phi$, 
and $n(x)$ denotes the particle's number density, reading  
\begin{equation}\label{scalarsource} 
n(x)=\int_{-\infty}^{\infty}\frac{1}{\sqrt{-g}}\hat{\delta}^4[x-z(\tau)]d\tau\,, 
\end{equation} 
where $g$ denotes the determinant of the background metric, and  
$\hat{\delta}^4[x-z(\tau)]$ formally represents a distribution which is similar  
to the four-dimensional Dirac delta-function in the sense that it vanishes 
away from the worldline. While the Dirac delta-function is an extremely useful 
mathematical tool to represent a structureless
 point particle in the context of a linear field equation,
 we find that it is somewhat less
suitable for the nonlinear field equation discussed in this 
paper. An alternative mathematical formulation which precisely encodes
the physical properties of the source is provided below.
 
We seek a perturbative solution to Eq. (\ref{scalarfeq}) 
in the form of    
\begin{equation}\label{scalpert} 
\phi=q\phi^{(1)}+q^2\phi^{(2)}+O(q^3)\,. 
\end{equation} 
In this expansion the perturbations $\phi^{(1)},\phi^{(2)}$ 
 are independent  
of $q$. 
Since $\phi$ is a dimensionless field, the charge $q$ 
has to form dimensionless combinations with other dimensional quantities
in each term in Eq. (\ref{scalpert}). The relevant dimensionless combinations are  
$q/\epsilon$ and $\{q/{\cal R}_i\}$, where $\{{\cal R}_i\}$ denotes the set
of length scales which characterize the background spacetime. 
Note that expansion (\ref{scalpert}) is valid only if these
 dimensionless quantities are kept small. This can be established by 
demanding that the charge of the particle will be 
much smaller than the length scales which characterize
the background geometry, 
i.e., $q\ll\cal{R}$, where ${\cal R}=\min\{{\cal R}_i\}$;
 and furthermore we have to assume that 
 $\epsilon\gg q$.
To satisfy the last inequality
 we shall use expansion (\ref{scalpert}) only in the external zone
defined to lie
outside a worldtube surrounding $z(\tau)$. We denote 
the interior of this worldtube by $S$ and define its surface $\Sigma$
 to lie at a spatial distance $\epsilon_E(q)$ from 
the worldline $z(\tau)$ such that $\epsilon_E\gg q$ (e.g., we may take $\epsilon_E=Nq$ where $N$ 
is some fixed large number).
Later we shall be interested in obtaining the limit $q\rightarrow 0$. In this 
limit the worldtube radius $\epsilon_E(q)$ approaches zero, and the perturbation analysis becomes valid throughout the entire background spacetime 
 excluding the worldline $z(\tau)$. 

We now seek a solution to Eq. (\ref{scalarfeq}) in the external zone.
For this purpose we substitute the perturbations series (\ref{scalpert}) 
 into Eq. (\ref{scalarfeq}) and obtain the following equations for the perturbations 
\begin{eqnarray}\label{scalar1st} 
&&\Box \phi^{(1)}=0 \ \ \ ,x\not\in S\\ \label{scalar2nd} 
&&\Box \phi^{(2)}=\nabla_{\mu}\phi^{(1)}\nabla^{\mu}\phi^{(1)} \ \ \ ,x\not\in S\,. 
\end{eqnarray} 
Notice that the form of these equations is independent of $q$.
As they stand Eqs. (\ref{scalar1st},\ref{scalar2nd}) contain no information 
about the physical properties of the  
charge which induces the scalar field. To obtain a unique solution
to each of these equations, we must provide additional information about this source.
Mathematically, this can be implemented by imposing a set of boundary conditions  
on $\Sigma$ which encode the physical properties of the source inside the worldtube
at each order.

Before we discuss these boundary conditions in detail, 
we briefly discuss their physical meaning, and provide a rough description of how
we intend to impose them mathematically. 
First, let us discuss the first-order equation (\ref{scalar1st}). 
To formulate the desired boundary conditions for this equation, we shall 
study the form of its general solution  
in a local neighborhood of $\Sigma$ within the external-zone.
Since $q\ll\cal{R}$, there exists a local neighborhood of $\Sigma$ in which
$\epsilon$ is much smaller than $\cal{R}$ and at the same time much larger than $q$; 
this region is called the buffer zone (e.g., in the buffer zone 
$\epsilon$ can be of order $\sqrt{q\cal{R}}$).  
By virtue of the smallness of $\epsilon/\cal{R}$ in the buffer zone,
 we may expand the general solution
to Eq. (\ref{scalar1st}), in powers of
${\cal R}^{-1}$, where the leading order in this expansion is simply 
the flat spacetime ``sum over multipoles'' solution. 
By assuming that the source of the scalar field
 is a structureless point particle, we 
find that its corresponding sum over multipoles 
can have no higher multipoles than a monopole.
The coefficient of this monopole term 
can later be determined from the value of the observed
 charge of the particle.
Below we show that by imposing a monopole solution  
on the boundary $\Sigma$ we obtain
a unique causal solution to Eq. (\ref{scalar1st})
in the limit $q\rightarrow 0$. 
Next we consider the second-order equation (\ref{scalar2nd}). A general solution 
of this inhomogeneous 
equation can be written as a sum of a particular inhomogeneous solution and 
a general homogeneous solution. To obtain a unique solution in the limit $q\rightarrow 0$,
 the freedom of adding an arbitrary homogeneous solution has to be constrained.
For this purpose we shall impose particular boundary conditions on the homogeneous solution.
Here again the absence of internal structure in the point-particle source 
implies that we should impose a monopole boundary condition on $\Sigma$.  

Below we shall provide more precise statements about the boundary conditions on $\Sigma$
 at orders $q$ and $q^2$, and we will 
show how they give rise to a unique causal solution to each of the equations 
 (\ref{scalar1st},\ref{scalar2nd}),
in the limit $q\rightarrow 0$.

\section{First-order perturbation}\label{firstorder}

Let us focus our attention on Eq. (\ref{scalar1st}).
Here we shall be interested in obtaining a unique causal solution 
in the limit $q\rightarrow 0$.

First, we shall identify the boundary conditions that 
have to be imposed on $\Sigma$ to obtain a unique causal solution to 
Eq. (\ref{scalar1st}) in the limit of interest. 
For this purpose, we shall express the solution to Eq. (\ref{scalar1st})
 using a Kirchhoff representation. In this way   
 the perturbation $\phi^{(1)}(x)$ can be expressed as a certain integral 
 over the worldtube surface $\Sigma$.
Assuming that the only source of the perturbation $\phi^{(1)}$ is the particle 
 itself (i.e. no incoming waves), and assuming that $\phi^{(1)}$ decays 
sufficiently fast at spatial infinity;
 we find that we may express $\phi^{(1)}$ as   
\begin{equation}\label{Kirchhoff} 
\phi^{(1)}(x)=-\frac{1}{4\pi}\int_{\Sigma(\epsilon_E)}\left[ 
G(x|x')\nabla^{\alpha'}\phi^{(1)}(x')-\phi^{(1)}(x')\nabla^{\alpha'}G(x|x') 
\right]d\Sigma_\alpha'\,. 
\end{equation} 
Here $G(x|x')$ denotes a Green's function (which is assumed to fall sufficiently fast into 
the past and the future), 
the surface integral is evaluated on the worldtube $\Sigma$, and  
$d{\Sigma'}_\alpha$ denotes the 
outward directed three-surface element on $\Sigma$. 
On physical grounds we shall require  
that $\phi^{(1)}$ will be a causal solution of Eq. (\ref{scalar1st}).  
To satisfy this requirement it is necessary to impose $ G(x|x')=G^{ret}(x|x')$, where
$G^{ret}(x|x')$ denotes the retarded scalar Green's function. 

Consider now the general solution $\phi^{(1)}$
in the buffer zone.
For the special case of a flat background spacetime, this  
general solution reduces to a standard sum over multipoles of a static
particle. Here the weight of each multipole corresponds    
to a particular moment of the particle's charge 
distribution. In curved spacetime, it is convenient 
to describe the background metric in a local 
neighborhood of the worldline which includes the buffer zone,
using a locally-flat coordinate system based
 on the worldline. Here we shall employ the Fermi normal coordinates.
In these coordinates the particle's  
geodesic trajectory is static. We can now 
solve Eq. (\ref{scalar1st})  
iteratively. This can be implemented by substituting 
 $g^{\mu\nu}=\eta^{\mu\nu}+\delta g^{\mu\nu}$ and
 $\phi^{(1)}=\phi^{(1)}_F+\delta \phi^{(1)}$ into Eq. (\ref{scalar1st}).
Here $\eta^{\mu\nu}$ is flat spacetime metric and $\phi^{(1)}_F$ is the corresponding 
general solution in flat spacetime. Notice that in the buffer-zone
both $\delta g^{\mu\nu}$
and $\delta \phi^{(1)}$ are small by virtue of the smallness
of $\epsilon/{\cal R}$ and $q/{\cal R}$. We assume that the internal structure 
of the point particle is unaffected by the background curvature.
In this case one finds that $\delta \phi(\epsilon_E)$ is always much smaller than
 $\phi^{(1)}_F(\epsilon_E)$ as $\epsilon_E \rightarrow 0$, and in fact 
$\lim_{\epsilon_E \rightarrow 0} \delta \phi(\epsilon_E)/\phi^{(1)}_F(\epsilon_E)=0 $.
We can formally express the desired general solution  
 on a family of hypersurfaces $\tau=const$, 
generated by spacelike geodesics with tangent vectors which are normal to  
worldline at $z(\tau)$, as
\begin{equation}\label{1stexpan} 
\phi^{(1)}(\tau,\epsilon,\Omega^\mu)=\sum_{m=1}^\infty a_{-m}\epsilon^{-m}+ 
\sum_{m=0}^{\infty}b_{m}\epsilon^{m}\,. 
\end{equation} 
Here the coefficients $\{a_{-m},b_m\}$ are independent of $\epsilon$, and  
 $\Omega^\mu$ denotes a unit vector which is normal to the worldline  
and tangent to the geodesic which connects $z(\tau)$ 
to $x$ on a particular hypersurface. 

By substituting the general solution (\ref{1stexpan}) on $\Sigma$ into Eq. (\ref{Kirchhoff}), 
and recalling that $d\Sigma_\alpha$ scales like ${\epsilon_E}^2$, 
 we find that the contribution to the integral which comes from
  the second sum in expansion (\ref{1stexpan}) vanishes in the limit $q\rightarrow 0$
(which implies $\epsilon_E(q) \rightarrow 0$), whereas the first sum in Eq. (\ref{1stexpan})
 which diverges as $x\rightarrow z(\tau)$, gives rise to a non-vanishing 
contribution to $\phi^{(1)}$.
Note that in the limit $q \rightarrow 0$
Eq. (\ref{scalar1st}) is valid throughout the entire spacetime
 excluding $z(\tau)$, reading
\begin{equation} \label{limscalar1st} 
\Box \phi^{(1)}=0 \ \ \ ,x\not\in z(\tau)\,.
\end{equation}
We conclude that it is sufficient to specify 
 {\em divergent boundary conditions}  to  
obtain a unique causal solution to Eq. (\ref{limscalar1st}).

To determine this solution we have to specify 
the coefficients $\{a_{-m}\}$ of the 
divergent terms in Eq. (\ref{1stexpan}).
First, let us consider the coefficients $\{a_{-m}\}$ with $m>1$. 
Recall that we assumed that the particle has no internal structure. 
This corresponds to a monopole solution in flat spacetime
 (i.e., $\phi^{(1)}_F=a_{-1}\epsilon^{-1}$).
Since the spacetime curvature induces corrections
of a smaller magnitude (as $\epsilon \rightarrow 0$), we find that 
 all the coefficients $\{a_{-m}\}$ with $m>1$ must vanish, and  
the coefficient $a_{-1}$ can have no angular dependence. 

To determine  $a_{-1}$ we consider the expansion of the full nonlinear field $\phi$
 in powers of $\epsilon$ in the buffer zone,
and identify the coefficient in front
 of the $1/\epsilon$ monopole term in this expansion to be the particle's observed charge.
In general this  $1/\epsilon$ term may depend on all the linear perturbations
 $\{\phi^{(i)}\}$.
For simplicity we shall assume that the observed charge is time-independent
and is identical to $q$. This implies that 
$a_{-1}=1$ and that 
the observed charge is completely determined by the first-order solution, 
without any contributions from higher-order perturbations. 

Fortunately there is a well-known retarded
solution to a linear homogeneous wave equation (\ref{limscalar1st}) 
with $1/\epsilon$ (monopole) singularity near the particle 
(see e.g., \cite{QUINN}), reading
\begin{equation}\label{phi1} 
\phi^{(1)}(x)=\int_{-\infty}^{\infty}G^{ret}[x|z(\tau)]d\tau\,. 
\end{equation} 
Since our divergent boundary conditions 
yield a unique causal solution, it must coincide with Eq. (\ref{phi1}). 

We comment that the same expression (\ref{phi1}) can be
explicitly derived by substituting the expansion (\ref{1stexpan}) with 
the desired coefficients into Eq. (\ref{Kirchhoff}), 
and only then taking the limit $\epsilon_E \rightarrow 0$. 
A similar method was used by D'Eath to obtain a solution to the first-order 
metric perturbations which are produced by a Schwarzschild black hole \cite{DEATH}. 
In this solution $h^{(1)}_{\alpha\beta}$ in the Lorenz gauge takes a form
 which is very similar to Eq. (\ref{phi1}). 

\section{Second-order perturbation}\label{secondorder}

We now focus our attention on Eq. (\ref{scalar2nd}).
In the limit $q\rightarrow 0$ this equation 
and its domain of validity take the following form
\begin{equation} \label{limscalar2nd} 
\Box \phi^{(2)}=\nabla_{\mu}\phi^{(1)}\nabla^{\mu}\phi^{(1)} \ \ \ ,x\not\in z(\tau)\,. 
\end{equation}
Notice that the form of the source term in this equation
 is similar to the form of the source term in the 
second-order gravitational perturbations equations (\ref{2ndgrav}); there the source term 
contains similar terms which are quadratic in  $\nabla_{\alpha}h^{(1)}_{\mu\nu}$. 

Naively one may try constructing the standard retarded solution to Eq. (\ref{limscalar2nd}) by 
integrating its source term with the retarded Green's function, reading  
\begin{equation}\label{divphi2} 
\phi^{(2)}_{ret}(x)=-\frac{1}{4\pi}\int G^{ret}[x|x']\nabla_{\mu'}\phi^{(1)}(x')\nabla^{\mu'}\phi^{(1)}(x') 
\sqrt{-g'} d^4x'\,. 
\end{equation} 
Here the integral extends over the entire spacetime. 
The difficulty with this naive approach is that the  
integral in Eq. (\ref{divphi2}) generically diverges for every point $x$ 
in spacetime. The origin of this divergency is the strong singularity of the   
source term  
 $\nabla_{\mu}\phi^{(1)}\nabla^{\mu}\phi^{(1)}$ near the particle. 
Recall that $\phi^{(1)}$ diverges 
like $1/\epsilon$, and therefore the second-order source term  
diverges like $\epsilon^{-4}$. Recalling that $\sqrt{-g'} d^4x'$ scales like $\epsilon 
^2$ we find that the expression inside the integral in Eq. (\ref{divphi2})  
 scales like $\epsilon^{-2}$, and therefore this integral diverges. 

Since the retarded solution (\ref{divphi2}) {\em diverges at every point in spacetime} 
it is not a legitimate solution to 
Eq. (\ref{limscalar2nd}). This difficulty might lead one to believe 
that  Eq. (\ref{limscalar2nd}) has no well defined (causal) solutions. 
Here we show that contrary to this belief there are  
in fact different causal solutions to equation (\ref{limscalar2nd}) which are 
 well defined, and have the desired physical properties. Notice that 
Eq. (\ref{limscalar2nd}) has the following particular solution 
\begin{equation}\label{psidef} 
\psi\equiv\frac{1}{2}\left[\phi^{(1)}\right]^2\,;
\end{equation} 
this can be easily verified by substituting $\psi$ into Eq. (\ref{limscalar2nd}).  
On physical grounds we have to make sure that $\psi$ is a causal solution.   
By a causal  
solution we mean: a solution to a field equation that at every point $x$ 
is completely determined by the given sources 
 (and/or boundary conditions) of this field 
equation that are inside and on the past lightcone 
associated with point $x$. 
Evidently, the retarded solution $\phi^{(1)}$ [given by Eq. (\ref{phi1})]
 is a causal solution of  
Eq. (\ref{limscalar1st}). Since $\psi$ is completely determined  
by $\phi^{(1)}$, we find that 
$\psi$ is a particular causal solution to Eq. (\ref{limscalar2nd}).  
 
We now discuss the general solution to Eq. (\ref{limscalar2nd}). 
Consider a different new solution to Eq. (\ref{limscalar2nd}), denoted with $\phi^{(2)}_N$. 
Defining the difference between $\phi^{(2)}_N$ and our   
original particular solution $\psi$ to be      
\[ 
\phi^{SH}\equiv\phi^{(2)}_N-\psi \,, 
\] 
we find that $\phi^{SH}$  
 satisfies the following semi-homogeneous equation 
 (i.e., a homogeneous equation away from the worldline) 
\begin{equation}\label{sheq} 
\Box \phi^{SH}=0 \,\,\, ,x\not\in z(\tau)\,. 
\end{equation} 
Clearly, the general solution to Eq. (\ref{limscalar2nd}) 
 that we denote $\phi^{(2)}_G$ can be written as $\phi^{(2)}_G=\psi+\phi^{SH}_G$, 
where $\phi^{SH}_G$ denotes the general solution to Eq. (\ref{sheq}). 
  
To construct a unique particular solution $\phi^{(2)}_P$ to Eq. (\ref{limscalar2nd}) 
we have to constrain the freedom associated with the 
arbitrariness of the semi-homogeneous part of $\phi^{(2)}_G$.
For this purpose we shall impose additional set of requirements on $\phi^{(2)}_G$. 
These requirements can then be transformed to a set of  
requirements on the semi-homogeneous field  
$\phi^{SH}_G$, yielding a    
unique particular solution to Eq. (\ref{sheq}), which we denote $\phi^{SH}_P$. 
By adding this particular solution to $\psi$ we shall obtain the desired particular 
 solution to Eq. (\ref{limscalar2nd}), $\phi^{(2)}_P$.  
 
Note that Eq. (\ref{sheq}) is identical to Eq. (\ref{limscalar1st}), and 
therefore it can be solved with a similar method.
Following our previous analysis of the first-order perturbation, we assume 
that the only source of (homogeneous) waves in 
the fields $\phi^{(2)}_P$ (and therefore in $\phi^{SH}_P$) is the particle 
 itself. Assuming that $\phi^{SH}_P$ decays 
sufficiently fast at spatial infinity we can express $\phi^{SH}_P$
using the Kirchhoff representation (\ref{Kirchhoff}).
 Following the first-order analysis 
we find that a unique causal solution $\phi^{SH}_P$ is completely
determined by its divergent boundary conditions as $x\rightarrow z(\tau)$.
Similar to Eq. (\ref{1stexpan}), we formally express $\phi^{SH}_P$ in the 
vicinity of $z(\tau)$ as 
\begin{equation}\label{shexpan} 
\phi^{SH}(\tau,\epsilon,\Omega^\mu)=\sum_{m=1}^\infty a^{SH}_{-m}\epsilon^{-m}+ 
\sum_{m=0}^{\infty}b^{SH}_{m}\epsilon^{m}\,. 
\end{equation} 
Here the divergent boundary conditions for the semi-homogeneous solution
$\phi^{SH}_P$ can be imposed by specifying 
the coefficients $\{a^{SH}_{-m}\}$ of the divergent terms in Eq. (\ref{shexpan}).

We now study the singular properties of $\phi$ in the buffer zone, 
and from them we shall obtain the required coefficients $\{a^{SH}_{-m}\}$.
For this purpose we consider $q$ to be a small (finite) quantity.
In the buffer-zone the first-order scalar perturbation  
$\phi^{(1)}(x)$ can be expanded 
 on a spacelike hypersurface that is generated by geodesics with tangent vectors
 which are normal to worldline at point $z(\tau_x)$, which gives
\begin{equation}\label{phi1dec} 
\phi^{(1)}(x)=\epsilon^{-1}+\phi^{(1)R}[z(\tau_x)]+O(\epsilon)\,. 
\end{equation} 
Here, $\phi^{(1)R}$ is a regular field 
 which solves the homogeneous 
scalar wave equation, generically this field scales like $\epsilon^{0}$ 
 (see \cite{DW} for the definition and properties of this field). 
Note that in Eq. (\ref{phi1dec}) the field  $\phi^{(1)R}$ is evaluated at   
 $z(\tau_x)$ on the worldline.  
By substituting Eq. (\ref{phi1dec}) into Eq. (\ref{psidef}) 
we obtain the following expansion 
\begin{equation}\label{phi2dec} 
\phi^{(2)}_P=\psi+\phi^{SH}_P=\frac{1}{2}\epsilon^{-2}+ 
\phi^{(1)R}[z(\tau_x)]\epsilon^{-1} 
+\phi^{SH}_P+O(\epsilon^0)\,. 
\end{equation} 
Substituting this expansion together with Eqs. (\ref{shexpan},\ref{phi1dec})  
into Eq. (\ref{scalpert}) gives 
\begin{equation}\label{phiexpan} 
\phi=\frac{q^2}{2\epsilon^{2}}+\epsilon^{-1} 
\Biglb[q+q^2(\phi^{(1)R}[z(\tau_x)]+a^{SH}_{-1})\Bigrb]+  
q^2\sum_{m=2}^\infty a^{SH}_{-m}\epsilon^{-m}+O(\epsilon^{0})+O(q^3)\,. 
\end{equation} 
As in the first-order analysis, the assumption
 that the particle has no internal structure implies that 
 the coefficients $ a^{SH}_{-m}$ with $m>1$ must vanish.
We now discuss the coefficient of the monopole $a_{-1}^{SH}$. 
Note that the second-term in Eq. (\ref{phiexpan}) has the form of a monopole term.  
We therefore identify the term inside the large square brackets with the observed scalar  
charge of the particle\footnote{We comment that unlike most linear field theories, the nonlinear field theory which is considered here does not have a unique 
definition for the particle's observed charge.} (up to second-order). Note that the term $\phi^{(1)R}[z(\tau_x)]$  
is time dependent, and might give rise to a time-dependent observed charge. 
Since we assumed that the particle's observed charge is time-independent and
 equals $q$, we find that   
\[ 
a^{SH}_{-1}=-\phi^{(1)R}[z(\tau_x)]\,. 
\] 
Having specified all the coefficients $\{a^{SH}_{-m}\}$ of the divergent terms
 in Eq. (\ref{shexpan}), the particular  
solution $\phi^{SH}_P$ is now uniquely determined. This solution takes the form  
\begin{equation} 
\phi^{SH}_P(x)=-\int_{-\infty}^{\infty}\phi^{(1)R}[z(\tau)]G^{ret}[x|z(\tau)]d\tau\,, 
\end{equation} 
as can be verified by an explicit calculation of the  
 $\{a^{SH}_{-m}\}$ coefficients of this solution. 
We conclude that the desired particular solution to Eq. (\ref{limscalar2nd}) is given by 
\begin{equation}\label{phi2}
\phi^{(2)}_P=\psi+\phi^{SH}_P=\frac{1}{2}[\phi^{(1)}]^2-\int_{-\infty}^{\infty}\phi^{(1)R}[z(\tau)]G^{ret}[x|z(\tau)]d\tau\,, 
\end{equation}
where $\phi^{(1)}$ is given by Eq. (\ref{phi1}). 
 
The derivation of the solution (\ref{phi2}) was based on the following steps:
(i) Considering the limit $q\rightarrow 0$ of the second-order perturbation equation
 (\ref{scalar2nd}).
 (ii) Guessing a particular inhomogeneous second-order solution $\psi$.
 (iii) Using the physical properties of the particle to construct divergent
 boundary conditions as $x\rightarrow z(\tau)$ for the semi-homogeneous field
$\phi^{SH}$. (iv) Solving the semi-homogeneous equation (\ref{sheq})
 with its divergent boundary conditions and obtaining $\phi^{SH}_P$. 

Finally we point out that the analysis given here can be generalized to 
handle the more problematic gravitational equation (\ref{2ndgrav}).
A detailed account on the regularization of the second-order gravitational 
perturbations will be published elsewhere.

\section*{Acknowledgments}

This research was carried out in collaboration with Amos Ori.
I am grateful to Amos for numerous valuable discussions.
I am grateful to Eric Poisson for valuable discussions, and for 
his comments on this manuscript. I thank the organizers of the seventh Capra meeting  
on radiation reaction for the generous hospitality. 

 
\end{document}